# Evaluation of CMIP models with IOMB: Rates of contemporary ocean carbon uptake linked with vertical temperature gradients and transport to the ocean interior

Weiwei Fu[1], J. Keith Moore[1], Francois Primeau[1], Nathan Collier[2], Oluwaseun O. Ogunro[2], Forrest M. Hoffman[2,3] and James T. Randerson[1]

[1] Department of Earth System Science, University of California, Irvine, CA 92697, USA

[2] Oak Ridge National Laboratory, Climate Change Science Institute (CCSI), Oak Ridge, TN 37831, USA

[3] Department of Civil & Environmental Engineering, University of Tennessee, Knoxville, TN 37996, USA

Date: 12/05/20

Submitted to:

Journal of Advances in Modeling the Earth System

Key points:

1) The International Ocean Model Benchmarking (IOMB) package is used to evaluate key upper ocean physical and biogeochemical variables from CMIP models.

2) Model performance overall improves from CMIP5 to CMIP6, particularly for surface nutrients, temperature, and salinity.

3) Biases in the anthropogenic ocean carbon sink are linearly related to biases in the vertical temperature gradient between 200 and 1000m for CMIP5 and CMIP6 models.

**Abstract:**

The International Ocean Model Benchmarking (IOMB) software package is a new community resource used here to evaluate surface and upper ocean variables from CMIP5 and CMIP6 Earth System Models (ESMs). IOMB generates charts and metrics of model performance for numerous variables describing ocean ecosystems, biogeochemistry, and physical properties, which enables comparison of the predictive skill of different models against observational datasets, and allows users to identify model deficiencies. Our analysis reveals general improvement in the multi-model mean of CMIP6 compared to CMIP5 for most of the variables we examined including surface nutrients, temperature, and salinity. We find that both CMIP5 and CMIP6 ocean models underestimate anthropogenic carbon dioxide uptake after the 1970s. For the period of 1994 to 2007, the multi-model mean from CMIP6 yields a mean cumulative carbon uptake of $27.2 \pm 2.2$ Pg C, which is about 15% lower than the $32.0 \pm 5.7$ Pg C estimate derived from two sets of observations. Negative biases in the change in anthropogenic carbon inventory exist in the northern North Atlantic and at mid-latitudes in the southern hemisphere (30-60°S). For the few models that provided simulations of chlorofluorocarbon (CFC), we demonstrate that regions with negative anthropogenic DIC biases coincide with regions that have a negative bias in CFC concentrations. This relationship suggests that underestimates of anthropogenic carbon storage in some models originates, in part, from weak transport between the surface and interior ocean. To examine the robustness of this attribution across the full suite of CMIP5 and CMIP6 models, we examined the vertical temperature gradient between 200 and 1000m as a metric for stratification and exchange between the surface and deeper waters. On a global scale across different models and different MIPs we find a linear relationship between the bias of vertical temperature gradients and the bias in anthropogenic carbon uptake, consistent with the hypothesis that model biases in the ocean carbon sink are related to biases in surface-to-interior transport. Our analysis shows that a low rate of $CO_2$ uptake is a common feature of most of the CMIP5 and CMIP6 models. The weak ocean $CO_2$ uptake implies some overestimation of the climate warming trend in model projections.

**Plain language summary**

With the increasing complexity of Earth System Models and more observations, we are motivated to develop an International Ocean Model Benchmarking system, which can quantitatively assess model performance and display diagnostics online. Novel charts of model performance were generated for numerous variables simulated by the models. Our analysis reveals general improvement in a newer generation of ocean biogeochemistry models used to support the 6[th] IPCC Assessment compared to an earlier generation of models used to support the 5[th] IPCC Assessment. Our analysis revealed that a common feature of both generations of ocean models is that they absorb less human-made carbon dioxide after the 1970s than what is indicated by two independent sets of observations. For the period of 1994-2007, the newer models absorbed, on average, 15% less carbon than what is indicated by the observations. Chlorofluorocarbon (CFC) simulations reported by a few of the models indicated that the low storage of fossil-derived carbon in the oceans results from weak transport between the surface and interior ocean. We examined the vertical temperature gradient as a metric for exchange between the surface and deeper waters. Globally, we found a strong linear relationship between bias of vertical temperature profiles and biases in the accumulation of fossil carbon, consistent with hypothesis that low rates of carbon uptake is caused in part by weak surface-to-interior transport. The weak ocean $CO_2$ uptake implies some overestimation of the climate warming trend in the model projections.

## 1 Introduction

The sixth phase of the Coupled Model Intercomparison Project (CMIP6) provides the science community with new climate models that incorporate many improvements and new features from different modelling groups (Aagaard & Carmack, 1989; Eyring et al., 2016; Gidden et al., 2019; O'Neill et al., 2016; Orr et al., 2017). The CMIP6 models are advancing our existing understanding of past, present and future climate. Ultimately, the best measure of the successes of the CMIP6 is the scientific progress on the most pressing problems of climate variability and climate projection (Eyring et al., 2016). A comprehensive validation of climate models constitutes a statistical foundation for climate studies because model bias and uncertainty have significant impacts on predicted climate variability, future projections and related downscaling and impact analysis (Flato et al., 2013). For CMIP6, comprehensive model validation can also support the IPCC Sixth Assessment Report (AR6) and other subsequent climate assessments reports at national and regional scales.

Among the processes needing validation, the uptake of carbon by the ocean is of paramount importance. A bias in carbon uptake by the oceans may lead to a bias in atmospheric $CO_2$ concentration in emissions-forced simulations, and consequently biases in the radiative forcing from anthropogenic emissions (Hoffman et al., 2014). For CMIP5, a comparison between the observed and modeled cumulative anthropogenic $CO_2$ uptake during the historical period was investigated by Bronselaer et al. (2017). They compared results of 14 models with estimates by Khatiwala et al. (2009) and found that 6 models were biased low and 6 models were biased high while 2 models agreed with observations for the period of 1791-1995. This analysis demonstrated the need to account for changes in the ocean inventory caused by anthropogenic forcing of atmospheric $CO_2$ prior to the start of CMIP5 simulations in 1850. Additional aspects of ocean carbon uptake were explored in a variety of studies linking carbon cycling with ocean ventilation and circulation (DeVries et al., 2017; Fletcher, 2017; Frolicher et al., 2015), carbon pumps (Yamamoto et al., 2018) and air-sea $CO_2$ exchange (Dong et al., 2016; Lauderdale et al., 2016). New estimates of anthropogenic carbon uptake derived from repeat ocean transects during 1974-2007 (Gruber et al., 2019) provide a new opportunity to evaluate recent changes in the ocean carbon cycle in earth system models (ESMs) from the 6[th] Phase of the Coupled Model

Intercomparison Project. These observations offer several advantages for model analysis compared to past efforts. First, corrections for anthropogenic $CO_2$ forcing prior to the start of the historical simulations in CMIP models (Bronselaer et al., 2017) are relatively small for this more recent period compared to adjustments required to assess changes in the full anthropogenic inventory from 1750 to the present, enabling a more robust comparison between the models and observations. Second, the new observations allow for an assessment of the models during a period of rapid atmospheric change, which may allow for evaluation of a different relative mix of processes and timescales regulating carbon uptake by the ESMs.

Here we conducted a detailed model evaluation for several important biogeochemical and physical climate variables for both CMIP5 and CMIP6 models using IOMB (the International Ocean Model Benchmarking (IOMB) software package) (Collier et al., 2018; Ogunro et al., 2018). Our first goal was to assess whether the representation of ocean biogeochemistry has improved from CMIP5 to CMIP6. The new IOMB package is able to make quantitative comparisons between time-dependent sequences of observed and simulated multidimensional fields of ocean biogeochemistry data, including sparsely distributed ocean interior observations, facilitating our analysis. Our second goal was to quantify model biases of anthropogenic ocean carbon uptake over the past several decades, and to investigate the relationship between these biases and physical processes.

## 2 Methods

### 2.1 CMIP5 and CMIP6 models

We compared biogeochemical variables from 11 CMIP5 and 9 CMIP6 Earth System Models (ESMs) with observations over the same time intervals using the IOMB software system. Table 1 provides a summary of the ocean and marine biogeochemical components of these models. These ESMs differ in the representation of ocean ecosystems and biogeochemistry, as well as the physical processes regulating ocean mixing and overturning. Most of the models revised their representation of marine biogeochemistry from CMIP5 to CMIP6 (Table 1). To evaluate model performance, we used 19[th] and 20[th] century simulations of climate change (referred to as the CMIP6 "historical" simulation) and a corresponding pre-industrial control simulation (referred to as the CMIP6 "piControl"). To obtain a multi-model mean state, we interpolated output fields from each model to standard grid with a 1° x 1° horizontal grid and a vertical resolution with 33 layers using linear interpolation. Note that the CMIP6 historical run spans the period from 1850–2014 while the CMIP5 historical simulation ends in 2005 with some models starting in 1860. The protocol for the historical simulation includes forcing by a common set of anthropogenic and natural driver variables derived from observations (Eyring et al., 2016). Both natural (e.g. solar variability and volcanic aerosols) and anthropogenic (e.g. greenhouse gas concentrations, aerosols, and land use) forcing influence climate variability and long-term trends in this simulation. The historical simulation protocol provides an effective means to compare model estimates with observations and to benchmark changes in model performance as individual models evolve over time. Differences in spin-up protocols are known to account for a substantial component of model disparities for biogeochemical fields ($O_2$, $NO_3$, alkalinity, dissolved Inorganic carbon), contributing to a relationship between spin-up duration and assessment metrics in CMIP5 models (Séférian et al., 2016). To remove the impact of model drift, here we used the difference between the historical and piControl simulations to estimate anthropogenic carbon uptake by the ocean.

## 2.1 Observational Datasets

We compared model simulations with a variety of observations from the 2018 release of the World Ocean Atlas (WOA18; (Boyer et al., 2019) and the Global Ocean Data Analysis Project Version 2 (GLODAPv2; (Olsen et al., 2016; Olsen et al., 2019) as well as SeaWIFS (Hu et al., 2012; NASA Goddard Space Flight Center et al., 2018) and MODIS-Aqua (NASA Goddard Space Flight Center et al., 2018) chlorophyll products. Mixed layer depths were evaluated against the product developed by de Boyer Montegut et al. (2004). Table 2 summarized key observations, their data sources, time span and corresponding model variables used here in IOMB.

A number of key variables from the models, including chlorophyll, nitrate, phosphate, oxygen, dissolved inorganic carbon (DIC), and alkalinity, were compared with observations at the surface in this configuration of IOMB because archived simulations from CMIP5 only recorded monthly output for these variables at the surface. Temperature and salinity were evaluated for surface, 200 m and 700 m levels for both CMIP5 and CMIP6 models because monthly subsurface fields were available. For ocean anthropogenic carbon inventory, we compared model output with two observationally-constrained estimates during 1994-2007, one by Gruber et al. (2019) drawing upon GLODAPv2 observations (hereafter GR2019) and a second estimate from DeVries (2014) based on inverse modelling using the GLODAPv1 data (hereafter DV2014). The DeVries (2014) product spans the 1780 to 2016 period; here we extracted observations from 1994-2014 to provide an independent assessment of recent changes in the ocean carbon inventory.

To facilitate the comparison, we binned the GLODAPv2 data into a standard three-dimensional grid, which had a 1° × 1° resolution in the horizontal and 33 levels vertically. The vertical layers in IOMB are identical to WOA standard levels, which have 12 uneven layers spanning 0-300m, 12 layers from 400 to 1500 m and 9 uneven layers from 1500 m to 5500 m. We performed this binning with a monthly resolution and obtained regridded GLODAPv2 data with linear interpolation. For the period of 1970-2010, for example, there are 963704 points for temperature and 159172 points for CFC11 in the 0-3000 m depth range, respectively.

## 2.2 IOMB

IOMB is a Python-based open-source, multi-model validation tool that can be used to evaluate the performance of CMIP5 and CMIP6 ocean biogeochemistry models. The IOMB package shares some code with its land model benchmarking counterpart, the International Land Model Benchmarking (ILAMB) package (Collier et al., 2018). The source code and documentation for IOMB can be found via https://www.ilamb.org/doc/running_iomb.html. The objectives of the IOMB project are: 1) to develop internationally accepted benchmarks for ocean model performance, 2) promote the use of these benchmarks by the international modelling community, 3) strengthen linkages between model and experimental communities in the design of new model tests and field measurement programs. The IOMB package has several features that make it effective for comparing across CMIP5 and CMIP6 model generations. First, IOMB can provide useful information about model performance taking into account for each variable the spatial pattern of annual mean bias and root mean square error (RMSE) as well as the phasing and amplitude of seasonal dynamics. Second, IOMB can simultaneously perform variable-to-variable comparison on multiple models for the same time period. Third, IOMB allows for the assessment of functional relationships between prognostic variables and one or more forcing variables. In past work, IOMB was used to benchmark aerosol precursors (Ogunro et al., 2018).

Against a seasonal climatology of monthly observations, we used IOMB to compare a number of diagnostic metrics including bias, RMSE, spatial distribution, and annual cycle, which are described in detail in Collier et al. (2018). IOMB provides model performance scores for each metric and generated a single scalar score for each variable by aggregating scores across metrics and datasets (Figure 1). IOMB has the capability to generate model comparisons at multiple depths. For the purpose of comparing CMIP5 and CMIP6 models, we only report comparisons here at the surface except for temperature and salinity. IOMB can generate a top-level, interactive summary page for each variable across models. The online version for the analysis reported here is available through https://www.ilamb.org/CMIP5v6/IOMB/dashboard.html. Clicking any square on the summary page for a particular variable, for example, allows the user to bring up new global maps comparing that model's mean output with the user-selected observational dataset (often a choice between GLODAPv2 and WOA2018) (Figure S1). Additional plots present global maps of the bias (the difference between model and observations) and root mean squared error (RMSE). Scores generated from these comparisons, along with information on seasonal phasing and interannual variability are integrated into a summary metric that is reported on the summary page. In addition, IOMB displays a table summarizing all the statistical metrics that go into the summary page for the selected variable, along with a Taylor diagram (Taylor et al., 2012) showing the overall fit for this variable for all the models (Figures S2-S3). Additional plots are generated for some variables relating variability and predictive skill over the annual cycle.

## 3 Results

### 3.1 CMIP5 and CMIP6 Model Evaluation with IOMB

Overall the representation of ocean biogeochemistry improved from CMIP5 to CMIP6 (Figure 1). The scalar scores of model performance are mapped in color, allowing users to visualize the improvement and quickly identify the relative performance of an individual model. The overall scores reported in Figure 1 integrate information on bias and RMSE as well as metrics that quantify differences between the models and the observations for the timing and phase of the annual cycle, interannual variability, and the spatial pattern of the annual mean field (Collier et al., 2018). To quantitatively assess, as whole, whether the CMIP6 models have improved, we report mean estimates from each MIP using IOMB in two different ways. First, we computed the mean of scores from individual models in each MIP (Table 3; first two columns).  Second, we created a mean map of each variable, by averaging together maps from individual models within each MIP. We constructed this multi-model mean by interpolating each individual model grid to a 1° x 1° horizontal resolution and to the WOA depth layers in the vertical dimension. We then ran IOMB on each of these mean fields, and report these scores in the final two columns of Table 2 and in the final two columns of Figure 1. Both approaches revealed that CMIP6 models had a higher overall score for 10 out of 13 variables. Notable exceptions where the models did not show improvement included surface chlorophyll, surface ALK, and temperature at 700m.

To further quantify model improvement, we report the bias and RMSE of the multi-model mean of the CMIP5 and CMIP6 models (Table 4).  The bias of multi-model mean was reduced by 20-70% for surface nitrate, phosphate, and silicate (Table 4) comparing the CMIP5 to CMIP6 models. Among them, the surface silicate showed most pronounced improvements. The bias of chlorophyll at the surface and of temperature at the surface and 700 m increased slightly from CMIP5 to CMIP6. For all of the variables we compared in IOMB, RMSE decreased in the CMIP6 models to varying degrees (Table 4). The RMSE of the multi-model mean was generally lower than the mean of individual model RMSE because across-model variability was reduced. In particular, surface DIC and total alkalinity were greatly improved. The DIC bias was decreased by 42% from 80 to 46 μmol/L and the RMSE was reduced by 46%. Similarly, the total alkalinity bias decreased by 45%, from 79 to 43 μmol/L. Note that the mixed layer depth bias was reduced to -7 m from 16 m in CMIP5, suggesting stronger stratification and weaker vertical transport in CMIP6 than in CMIP5.  The full suite of linked graphics and statistical metrics cannot be shown in this paper, but are available to readers via the following link https://www.ilamb.org/CMIP5v6/IOMB/dashboard.html.

While the multi-model means indicated general improvement, there was no consistent improvement pattern across models and variables. For example, surface silicate was improved in almost all the models, but the improvements were more noteworthy in the HadGEM2 and GFDL models. Compared with the WOA18 data, the bias was reduced from 45.5 to 2.5 mmol/m³ from CMIP5 to CMIP6 for HadGEM2 while the bias was reduced from 8.4 to 1.0 mmol/m³ in GFDL-ESM2G. In CMIP6, the surface silicate of CNRM-ESM2 had the lowest bias of -0.4 mmol/m$^3$ followed by GFDL-ESM4. The surface silicate of MPI- MPI-ESM1-2-HR showed the highest bias of 9.1 mmol/m$^3$. For surface chlorophyll in CMIP5, CESM1-BGC gave the highest score followed by the IPSL models, while CanESM2 had the lowest score. However, a model with a relatively high score in CMIP5 did not guarantee a similar ranking for the successor model. In CMIP6, the prediction skill of CESM was degraded for surface chlorophyl while the GFDL-CM4 model was significantly improved and showed the highest score. The score chart in Figure 1 also calls attention to inconsistent improvements for different variables such as nitrate, phosphate and silicate within a single model. The model with a high score for one macronutrient may have a poor rating for another, and vice versa. A detailed investigation of these differences is outside the scope of this paper, but further IOMB analysis may help clarify covariances and interactions among some of the drivers.

## 3.2 Anthropogenic ocean carbon uptake

For the 1994 to 2007 period, we compared anthropogenic carbon $C_{ant}$ storage estimates from the CMIP6 models with data-derived estimates from DV2014 and GR2019 (Figure 2, Table 5). In the context of this comparison, it is important to note that the CMIP6 model historical simulations begin from a steady state in the year 1850, and miss ocean carbon accumulation associated with pre-1850 anthropogenic $CO_2$ increases. Therefore, we adjusted the models to account for the pre-1850 $CO_2$ forcing. According to Bronselaer et al. (2017), the adjustment corresponding to the 1994-2007 period was about 0.7 Pg C, corresponding to about a 2.4% positive increase relative to the total accumulation during this interval reported by DV2014. After this adjustment, the CMIP6 multi-model mean value for $C_{ant}$ was 27.2 ± 2.2 Pg C. In contrast, observationally constrained estimates for this period were 33.7 ± 4.0 Pg C from GR2019 and 30.2 Pg C from DV2014. To create a mean estimate for the observations, we averaged together the GR2019 and DV2014 estimates. We created an uncertainty range by assuming the DV2014 estimate had a similar level of uncertainty as the GR2019 analysis, recognizing these studies employ fundamentally different methods. We then combined the uncertainty estimates from the two studies together in quadrature. This yielded a mean observed estimate of 31.9 ±5.7 Pg C. Relative to this mean, the CMIP6 multi-model mean was lower by 15%. 8 of 9 models from CMIP6 had $C_{ant}$ estimates below the mean of the observations. Only GFDL-CM4 model simulated $C_{ant}$ storage at a level above the mean. The low model bias varied for different ocean basins. The multi-model mean underestimated $C_{ant}$ storage by 2.2 Pg C in the Atlantic Ocean (19%) and by 1.4 Pg C in the Pacific as compared to mean of the observations. The Atlantic Ocean accounted for about 45% of the low bias in global $C_{ant}$ storage. For the CMIP5 models, similarly, the multi-model mean of $C_{ant}$ was 3.1 Pg C lower than DV2014 for the period from 1994 to 2005. Only the NorESM1-ME model simulated a $C_{ant}$ close to that of DV2014 among the 11 models. The Atlantic Ocean was also the location of the largest low bias (among other ocean basins) in the CMIP5 models.

For the observations, the maximum uptake and storage of anthropogenic $CO_2$ was in the high-latitude North Atlantic and at mid-latitudes across the Southern Hemisphere (Figure 3). Most of the models capture some enhanced uptake and storage in the Southern Ocean, but struggle to reproduce the Southern Hemisphere maximum in storage associated with the formation of Subantarctic Mode and Antarctic Intermediate Waters (Frolicher et al., 2015). The formation of these water masses is key to moving anthropogenic $CO_2$ from the surface to the ocean interior. The two GFDL models did capture the strong uptake and storage in mid-latitudes of the Southern Hemisphere, but most models did not, suggesting weak formation of these intermediate water currents, at least in some ocean basins (Figure 3). A number of the models show hot spots of very strong $C_{ant}$ uptake in the high latitude Southern Ocean (south of 60°S), likely as a consequence of open-ocean deep convection, a common problem in coarser resolution ESM ocean models (Canuto et al., 2004; Reintges et al., 2017). The Southern Ocean south of 40°S accounts for 35% of the global anthropogenic $CO_2$ uptake from the atmosphere from 1994 to 2007. Upwelling at the Antarctic Divergence brings deep water with very low anthropogenic $CO_2$ to the surface, facilitating uptake of a large amount of $C_{ant}$ in the presence of high winds.

The oceans stored less anthropogenic carbon in the tropics, the northern Indian Ocean and in the northern Pacific Ocean. The low storage in these regions results from the large transport of anthropogenic carbon out of these regions and higher levels of stratification compared to other ocean regions (Frolicher et al., 2015). Some CMIP6 models also simulated large biases in carbon storage in the northern tropical Pacific and the Kuroshio Extension (Figure 3).

### 3.3 Biases of Anthropogenic $CO_2$, CFCs and Vertical Temperature Gradients

The low bias of $C_{ant}$ in recent decades may be attributable to both physical and biogeochemical processes. To assess biases in ocean circulation and mixing, comparison of simulated CFC distributions from CMIP6 models with observations offers the possibility of directly assessing the magnitude of exchange between surface and sub-surface waters. Because of the known time history of atmospheric concentrations, and the fact that CFCs are biologically inert and very stable in the ocean, they serve as unambiguous tracers of ocean circulation (Dutay et al., 2002; England et al., 1994). By comparison with GLODAPv2 CFC observations, we found that the four CMIP6 models that reported CFC estimates, all had a low bias in the global ocean inventory over the period of 1994-2007. This suggested that vertical mixing and transport to the interior in the models was too weak. Further analysis revealed that the spatial structure of CFC and anthropogenic DIC biases in the model were correlated (Figure 4). The joint density distribution showed that a low bias of model anthropogenic DIC coincides with a low bias of model CFC11, which indicates weak transport from surface to the ocean interior. For GFDL-CM4 model, positive CFC biases occurred at more grid points than for the CESM2 and CESM2-WACCM, consistent with the higher $C_{ant}$ uptake by GFDL-CM4 (Figures 3 and 4). Particularly in the Southern Hemisphere GFDL-CM4 had more positive CFC biases and larger $C_{ant}$ storage than the other models.

To examine the robustness of the relationship between CFC bias and $C_{ant}$ bias, we need more CFC output from CMIP6 models. Unfortunately, most model centers have not yet uploaded their CFC output to the Earth System Grid (https://esgf-node.llnl.gov/search/cmip6/). Moreover, CFCs were not a standard output in CMIP5. As another tracer of vertical ocean mixing, we compared simulated and observed vertical temperature gradients between 200m and 1000m. This tracer of mixing has several advantages. First, all of the CMIP models report the three-dimensional structure of ocean temperature. Second, in classical models of ocean circulation, vertical temperature gradients and vertical mixing together determine ocean heat uptake (Munk, 1966). In this paradigm, downward heat flux by small-scale mixing balances a thermally directed overturning circulation that transports heat upward. Therefore, the transport of heat and other tracers from the surface to the interior is expected to be closely related to strength of the vertical temperature gradients.

We computed the temperature gradient (dT/dZ) for each model grid location by fitting a linear equation to vertical temperature profile from 200 m to 1000 m where existed most $C_{ant}$ uptake. The dT/dZ was similarly computed for the WOA18 and GLODAP v2 data, which was averaged to a 1° x 1° grid. The profiles in Figure. S5 show the global mean profiles for the models and the observations. The bias of dT/dZ for different models is shown in Figure. 5 during the period of 1994-2007. Most of models had a positive dT/dZ bias in the Southern Ocean, which was consistent with the lower DIC inventory in this region. The bias of dT/dZ by comparison with the WOA18 data showed similar spatial patterns to the GLODAPv2 data in different basins (the figures are available on the online IOMB version). The bias of dT/dZ in GFDL-CM4 and CNRM-ESM2 models was the smallest among the models for the 30°S-50°S latitude belt, which corresponded to strong DIC uptake in these models. In other regions such as the North Atlantic and North Pacific, we also noticed relationships between the bias of dT/dZ and biases in the DIC inventory. We related the dT/dZ bias to the CFC bias by calculating the joint density between them in Figure. 6. Strong relationships between these biases were found in the four models that we previously examined in Figure 4, which had significant negative correlations between CFCs and $C_{ant}$. This is reasonable because strong stratification inhibits vertical downward transport of heat and other tracers like $C_{ant}$.

Scatter plots of anthropogenic DIC bias and dT/dZ bias demonstrated a robust, negative relationship across the 9 CMIP6 models (Figure. 7). For the CMIP5 models, we also calculated the bias of temperature gradient and $C_{ant}$ storage change during the period of 1994-2005 (Fig. S6 and Table S1). The biases of $C_{ant}$ storage during this period from DV2014 were also strongly correlated with the bias in dT/dZ.

There was a strong correlation between the dT/dZ bias and the bias of DIC inventory across the CMIP5 and CMIP6 models (Figure. 8). We fit a linear function between the bias of DIC inventory and vertical temperature gradient, which has a form of $Y = -3.5 \frac{Pg\,C}{°C/Km}\,X - 2.8$ on the global scale. Here, Y is the bias of DIC inventory (Pg C) and X is the bias of vertical temperature gradient (°C/Km). This negative relationship explained about 70% of model-to-model differences in $C_{ant}$ biases.. The weak downward transport, especially in the mid-latitudes of the southern hemisphere, inhibits the transport of anthropogenic $CO_2$ via the formation of intermediate and Subantarctic mode waters. This was consistent with the large negative bias of $C_{ant}$ storage from 30-60°S (Figure. 3). Specifically, the IPSL model with the largest positive bias also showed the largest negative $C_{ant}$ bias at these latitudes while GFDL models showed positive $C_{ant}$ biases.

## 4 Discussion and Conclusions

Using the IOMB package, we assessed the performance of ocean ecosystem and biogeochemistry models from CMIP5 and CMIP6. A number of physical and biogeochemical variables were compared with observations at the surface, 200 m and 700 m levels. In contrast to other diagnostic tools, IOMB generates a score table of model performance by aggregating different metrics, allowing for a quantitative estimate of model prediction skill for a specific variable at a given depth. We found that the performance of CMIP6 models was generally better than the CMIP5 models with bias and RMSE reduced for most model variables, but the extent of improvement varied depending on variable and individual model. Overall scores improved in CMIP6 for 11 of 14 variables, with exceptions for surface chlorophyll, temperature at 700m, and salinity at 700 m. These latter variables showed small levels of degradation in the multi-model mean. Overall, the summary chart of the IOMB is able to provide useful information for future studies of the CMIP5 and CMIP6 models, and complements the ILAMB package, which evaluates terrestrial physical and biogeochemical variables against observational datasets (Collier et al., 2018).

We estimated anthropogenic ocean carbon storage rate and cumulative carbon in the historical simulations over the period of 1994-2007. The CMIP6 models predicted a a multi-model mean of 27.2±2.2 Pg C , which was considerably lower than the estimate of 33.7±4.0 Pg C by Gruber et al. (2019) using GLODAPv2 data and the estimate of 30.2 Pg C using inverse methods by DeVries (2014). Large negative biases existed the northern North Atlantic Ocean and the mid-latitudes of the Southern Hemisphere (30-60°S). Similarly, the anthropogenic carbon storage of CMIP5 models was also lower than DV2014 estimate during the period of 1994-2005. Only the NorESM1-ME out of 11 CMIP5 models had higher than observed ocean carbon storage.

We also found a negative model bias for the CFCs, which was strongly correlated with the bias for anthropogenic DIC in CMIP6 models. As CFCs were not available for all models, we also examined the strength of vertical temperature gradients, which were available for most of the CMIP5 and CMIP6 models. Across the 16 CMIP5 and CMIP6 models, a positive vertical temperature gradient bias coincided with negative Cant storage bias, providing additional evidence that weak ocean transport results in less anthropogenic carbon uptake.

Comparison of multiple models with a top-level score chart is a great advantage of the IOMB over more traditional, single-model, diagnostic tools. It is important to recognize that the single score summarizing model performance is derived using our choice of metrics, which issubjective as in other evaluation tools. This is important to consider when interpreting the scores. In addition to the metrics used in the current version, IOMB can be expanded by incorporating other benchmarking datasets and metrics from the ocean community. The score chart in Figure 1 can be considered an initial, useful evaluation of the CMIP5 and CMIP6 models.

We evaluated the CMIP6 models at three different depth levels with IOMB. Ongoing IOMB development efforts include adding more model output and observational datasets, adding more types of plots for each variable (comparing with different observational transects, along isopycnal layers and sub-setting by different ocean basins and/or biomes). We are also adding new features to IOMB for a better assessment of the long-term response of the ocean's carbon cycle to global warming. We expect IOMB to produce a comprehensive evaluation of ocean model performance, to help model developers identify deficiencies and subsequently accelerate model development, and to facilitate non-specialist routine analysis for climate and ocean studies.

We quantified $C_{ant}$ biases and connected them with biases in ocean vertical transport. However, anthropogenic ocean carbon uptake is influenced by many processes, including, for example,partial pressure differences, solubility, circulation, and the strength of thebiological pump. These processes are related to each other, which compounds the challenge of attribution of the $C_{ant}$ storage bias. The vertical temperature gradient we examined here is associated the meridional overturning circulation in the North Atlantic and deep convections in the Southern Ocean. Solubility effects also lead to a positive feedback, which is however of secondary importance compared with ocean dynamics (Crueger et al., 2008). As shown in Marinov and Gnanadesikan (2011), air-sea $CO_2$ flux is not sensitive to ocean circulation but the storage of ocean carbon is sensitive to ocean circulation, which redistributes the uptake of $C_{ant}$ in the global ocean. The relationship between the bias of anthropogenic ocean DIC and vertical temperature gradient requires further exploration with other diagnostics of circulation and mixing, but the attribution seems robust and the bias in ocean transport appears to be of first order importance in regulating model-to-model differences storage of anthropogenic carbon in the oceans during the historical period.

### Acknowledgements

The authors acknowledge support from the Reducing Uncertainty in Biogeochemical Interactions through Synthesis and Computation (RUBISCO) Scientific Focus Area (SFA), which is sponsored by the Regional and Global Model Analysis (RGMA) program area of the Earth & Environmental Systems Sciences Division (EESSD) of the Biological and Environmental Research (BER) office of the U.S. Department of Energy (DOE) Office of Science. We acknowledge the World Climate Research Programme's Working Group on Coupled Modelling, which is responsible for CMIP. We thank the climate modeling groups for producing and making available their model output, the Earth System Grid Federation (ESGF) for archiving the data and providing access, and the multiple funding agencies who support CMIP5, CMIP6, and ESGF. We thank DOE's RGMA program area, the Data Management Program, and NERSC for making this coordinated CMIP6 analysis activity possible. The CMIP5 and CMIP6 data can be accessed using the link https://esgf-node.llnl.gov/search/cmip5/ and https://esgf-node.llnl.gov/search/cmip6/.

Table 1. The CMIP5 and CMIP6 ocean models evaluated here using IOMB.

| CMIP5 | | | CMIP6 | | |
|---|---|---|---|---|---|
| ESM | Ocean | Ocen BGC | ESM | Ocean | Ocean BGC |
| MPI-ESM-LR (Giorgetta et al., 2013) | MPI-OM (1°×1.4°) | HAMOCC v5.2 (Ilyina et al., 2013) | MPI-ESM1-2-LR (Muller et al., 2018) | MPI-OM (1.5°×1.5°) | HAMOCC6 (Paulsen et al., 2017) |
| MPI-ESM-MR (Giorgetta et al., 2013) | MPI-OM (1.41°×0.89°) | HAMOCC v5.2 (Ilyina et al., 2013) | MPI-ESM1-2-HR (Muller et al., 2018) | MPI-OM (0.4°×0.4°) | HAMOCC6 (Paulsen et al., 2017) |
| IPSL-CM5A-LR (Dufresne et al., 2013) | NEMO-ORCA2 (2°×2°) | PISCES (Aumont & Bopp, 2006) | IPSL-CM6A-LR (Boucher et al., 2020) | NEMO-eORCA1 (1°×1-1/3°) | PISCES v2 (Aumont et al., 2015) |
| HadGEM2-ES (Collins et al., 2011; Jones et al., 2011) | HadGOM2 (0.3-1°×1°) | Diat-HadOCC (Totterdell, 2019) | UKESM1 (Sellar et al., 2019) | NEMO-ORCA2 (2°×2°) | MEDUSA2 (Yool et al., 2013) |
| CESM1(BGC) (Gent et al., 2011; Hurrell et al., 2013) | POP2 (1°×1°) | BEC (Moore et al., 2004; Moore et al., 2013) | CESM2 (Danabasoglu et al., 2020) | POP2 (1°×1°) | BEC (Moore et al., 2004; Moore et al., 2013) |
| NorESM1-ME (Bentsen et al., 2013) | MICOM (1.125°) | HAMOCCv5.1 (Assmann et al., 2010; Ilyina et al., 2013) | NorESM2 (Seland et al, 2020) | MICOM-Tripolar (0.5°×0.9°) | HAMOCCv5.1 (Schwinger et al., 2016; Tjiputra et al., 2013) |
| CNRM-CM5 (Voldoire et al., 2013) | NEMO-ORCA1 (1°×1°) | PISCES (Aumont et al., 2003) | CNRM-ESM2-1 (Seferian et al., 2019) | NEMO-eORCA1 (1°×1°) | PISCES v2 (Aumont et al., 2015) |
| CanESM2 (Arora et al., 2013) | CanOM4 (0.9°×1.4°) | CMOC (Christian et al., 2010; Zahariev et al., 2008) | CanESM5 (Swart et al., 2019) | NEMO-ORCA1 (1°×1-1/3°) | CMOC (Christian et al., 2010; Zahariev et al., 2008) |
| GFDL-ESM2G (Dunne et al., 2012; Dunne et al., 2013) | TOPAZ-Tripolar (1°×1°) | TOPAZ2 (Dunne et al., 2013) | GFDL-ESM4 (Dunne et al., 2020) | MOM6 (0.5°) | COBALTv2 (Stock et al., 2014) |
| GFDL-ESM2M (Dunne et al., 2012; Dunne et al., 2013) | MOM4-Tripolar (1°×1°) | TOPAZ2 (Dunne et al., 2013) | GFDL-CM4 (Held et al., 2019) | MOM6 (0.25°) | BLINGv2 (Dunne, Bociu, et al., 2020) |

Table 2. Data products for different biogeochemical and physical variables integrated within IOMB.

| Observations | Model variable | Data sources and references | Temporal coverage |
|---|---|---|---|
| Chlorophyll-a | chl | GLODAPv2 (Key et al., 2015; Olsen et al., 2016) | 1970-2010 |
| | | SeaWIFS (Hu et al., 2012; NASA Goddard Space Flight Center et al., 2018) | 1997-2010 |
| | | MODIS-Aqua (NASA Goddard Space Flight Center et al., 2018) | 1997-2010 |
| Oxygen | o2 | GLODAPv2 (Key et al., 2015; Olsen et al., 2016) | 1970-2010 |
| | | WOA2018 (Garcia et al., 2019a) | 1955-2010 |
| Nitrate | no3 | GLODAPv2 (Key et al., 2015; Olsen et al., 2016) | 1970-2010 |
| | | WOA2018 (Garcia et al., 2019b) | 1955-2010 |
| Phosphate | po4 | GLODAPv2 (Key et al., 2015; Olsen et al., 2016) | 1970-2010 |
| | | WOA2018 (Garcia et al., 2019b) | 1955-2010 |
| Silicate | si | GLODAPv2 (Key et al., 2015; Olsen et al., 2016) | 1970-2010 |
| | | WOA2018 (Garcia et al., 2019b) | 1955-2010 |
| Dissolved Inorganic Carbon | dissic | GLODAPv2 (Key et al., 2015; Olsen et al., 2016) | 1970-2010 |
| | | OCIM (DeVries, 2014) | 1994-2007 |
| | | Gruber (Gruber et al., 2019) | 1994-2007 |
| Alkalinity | alk | GLODAPv2 (Key et al., 2015; Olsen et al., 2016) | 1970-2010 |
| Temperature | thetao | GLODAPv2 (Key et al., 2015; Olsen et al., 2016) | 1970-2010 |
| | | WOA2018 (Locarnini et al., 2019) | 1955-2010 |
| | | LDEO (Reynolds et al., 2002) | 1955-2010 |
| Salinity | so | GLODAPv2 ((Key et al., 2015; Olsen et al., 2016) | 1970-2010 |
| | | WOA2018 (Zweng et al., 2019) | 1955-2010 |
| Mixed layer depth | mlotst | de Boyer Montegut et al. (2004) | 1941-2002 |

Table 3. Overall score of model performance for different variables in CMIP5 and CMIP6 models. The overall score is calculated based on bias, RMSE, phase shift, interannual variability and spatial distribution. The mean of scores from individual models listed in Table 1 are shown in the first two columns, while the application of IOMB to a single mean field constructed from all of the CMIP5 or CMIP6 models is shown in the final two columns.

| Variables | Mean of scores from individual CMIP5 models | Mean of scores from individual CMIP6 models | Score derived from the CMIP5 mean field | Score derived from the CMIP6 mean field |
|---|---|---|---|---|
| Chlorophyll at surface | 0.344 | 0.341 | 0.325 | 0.322 |
| Nitrate surface | 0.462 | 0.465 | 0.455 | 0.472 |
| Phosphate surface | 0.511 | 0.523 | 0.538 | 0.554 |
| Silicate surface | 0.421 | 0.462 | 0.429 | 0.500 |
| Oxygen surface | 0.528 | 0.552 | 0.560 | 0.578 |
| Temperature surface | 0.602 | 0.613 | 0.641 | 0.647 |
| Temperature 200m | 0.468 | 0.481 | 0.509 | 0.517 |
| Temperature 700m | 0.463 | 0.461 | 0.492 | 0.487 |
| Salinity surface | 0.471 | 0.483 | 0.502 | 0.516 |
| Salinity 200m | 0.468 | 0.473 | 0.498 | 0.510 |
| Salinity 700m | 0.461 | 0.462 | 0.498 | 0.499 |
| TALK surface | 0.364 | 0.362 | 0.394 | 0.378 |
| DIC surface | 0.368 | 0.392 | 0.389 | 0.392 |
| Mixed Layer Depth | 0.501 | 0.562 | 0.512 | 0.634 |

Table 4. Mean bias and RMSE of individual CMIP5 and CMIP6 models. The bias and RMSE of multi-model mean variables is also given in the last two columns. The multi-model mean was calculated by interpolating original model grid to 1x1° grid in the horizontal.

| Variables | Mean of CMIP5 | | Mean of CMIP6 | | CMIP5 mean | | CMIP6 mean | |
|---|---|---|---|---|---|---|---|---|
| | Bias | RMSE | Bias | RMSE | Bias | RMSE | Bias | RMSE |
| Chlorophyll at surface ($mg/m^3$) | -0.04 | 0.54 | -0.06 | 0.45 | -0.04 | 0.55 | -0.06 | 0.38 |
| Nitrate surface ($\mu M$) | 0.84 | 2.59 | 0.67 | 2.29 | 0.84 | 1.99 | 0.67 | 1.66 |
| Phosphate surface ($\mu M$) | -0.08 | 0.21 | -0.02 | 0.20 | -0.12 | 0.18 | -0.06 | 0.16 |
| Silicate surface ($\mu M$) | 8.10 | 10.51 | 2.21 | 5.52 | 8.10 | 9.21 | 2.21 | 4.44 |
| Oxygen surface ($\mu M$) | 6.38 | 12.55 | 3.27 | 10.49 | 6.38 | 11.21 | 3.27 | 8.79 |
| Temperature surface (°C) | -0.45 | 1.53 | -0.48 | 1.36 | -0.45 | 1.07 | -0.48 | 1.00 |
| Temperature 200m (°C) | 0.05 | 1.69 | 0. 04 | 1.52 | 0.05 | 1.18 | 0. 04 | 1.09 |
| Temperature 700m (°C) | 0.49 | 1.32 | 0.50 | 1.29 | 0.49 | 1.03 | 0.51 | 0.97 |
| Salinity surface (PSU) | -0.23 | 0.71 | -0.08 | 0.64 | -0.23 | 0.57 | -0.08 | 0.47 |
| Salinity 200m (PSU) | -0.18 | 0.37 | -0.09 | 0.34 | -0.18 | 0.27 | -0.09 | 0.22 |
| Salinity 700m (PSU) | 0.03 | 0.23 | 0.03 | 0.22 | 0.03 | 0.14 | 0.03 | 0.14 |
| TAlk surface ($\mu mol/L$) | 78.9 | 99.7 | 43.2 | 64.0 | 78.9 | 99.5 | 43.2 | 47.8 |
| DIC surface ($\mu mol/L$) | 80.1 | 89.9 | 45.8 | 57.8 | 80.12 | 90.8 | 45.8 | 48.3 |
| Mixed Layer Depth (m) | 16.1 | 46.9 | -7.4 | 34.5 | 16.1 | 40.5 | -7.4 | 26.4 |

Table 5. Change of anthropogenic DIC inventory (Pg C) in the layer of 0-3000 m for different CMIP6 models is compared with the DV2014 estimate and GR2019 data from 1994 to 2007. The DV2014 data starts in the year 1780. Adjustment was made to account for the impact of pre-1850 CO2 rise, which leads to the numbers in parenthesis. The Arctic and other marginal seas are not included. The numbers for GR2019 data can be found in table S3 of Gruber et al. (2019). The Atlantic Ocean, Indian Ocean, and Pacific Ocean regions in the first three columns extend to Antarctica to enable direct comparison with observational estimates from Gruber et al. (2019). These three columns sum to the Global Ocean estimate. To assess biases over the Southern Ocean we separately report OCIM and model carbon inventories south of 40°S in the final column.

| | Change of anthropogenic DIC inventory from 1994 to 2007 (Pg C) | | | | |
| | Global ocean | Atlantic Ocean | Indian Ocean | Pacific Ocean | Southern Ocean (South of 40°S) |
|---|---|---|---|---|---|
| **Gruber et al. 2019** | **33.7±4.0** | **11.9±1.3** | **7.1±3.4** | **13.2±1.3** | **NA** |
| **DeVries 2014 (OCIM)** | **30.2** | **11.0** | **6.5** | **12.7** | **5.7** |
| **Mean of obs.** | **32.0±5.7** | **11.5±1.8** | **6.8±4.8** | **13.0±1.8** | **5.7** |
| **Multi-model mean (std)** | **27.2±2.2** | **9.3±1.8** | **6.3±1.1** | **11.6±1.0** | **5.4±1.4** |
| CESM2 | 26.1 | 9.7 | 6.7 | 9.7 | 6.1 |
| CESM2-WACCM | 25.9 | 9.7 | 5.6 | 10.6 | 5.1 |
| CanESM5 | 27.0 | 8.7 | 6.1 | 12.2 | 5.7 |
| CNRM-ESM2 | 27.4 | 8.4 | 6.4 | 12.7 | 4.7 |
| GFDL-CM4 | 32.1 | 13.3 | 5.8 | 13.0 | 4.1 |
| GFDL-ESM4 | 29.1 | 8.2 | 9.1 | 11.8 | 7.7 |
| IPSL-CM6A-LR | 24.3 | 6.6 | 5.3 | 12.4 | 2.9 |
| MPI-ESM1-2-HR | 25.6 | 8.9 | 5.0 | 11.7 | 4.7 |
| NorESM2-LM | 27.3 | 10.3 | 6.2 | 10.8 | 7.4 |